\documentclass[12pt]{article}
\input epsf
\makeatletter

\def\newpic#1{}

\def\hybrid{\topmargin 0pt      \oddsidemargin 0pt
             \headheight 0pt \headsep 0pt

           \textwidth 6.25in       
             \textheight 9.5in       
             \marginparwidth 0.0in
             \parskip 5pt plus 1pt   \jot = 1.5ex}
\catcode`\@=11
\def\marginnote#1{}

\newcount\hour
\newcount\minute
\newtoks\amorpm
\hour=\time\divide\hour by60
\minute=\time{\multiply\hour by60 \global\advance\minute by-\hour}
\edef\standardtime{{\ifnum\hour<12 \global\amorpm={am}%
             \else\global\amorpm={pm}\advance\hour by-12 \fi
             \ifnum\hour=0 \hour=12 \fi
             \number\hour:\ifnum\minute<10 0\fi\number\minute\the\amorpm}}
\edef\militarytime{\number\hour:\ifnum\minute<10 0\fi\number\minute}

\def\draftlabel#1{{\@bsphack\if@filesw {\let\thepage\relax
        \xdef\@gtempa{\write\@auxout{\string
           \newlabel{#1}{{\@currentlabel}{\thepage}}}}}\@gtempa
        \if@nobreak \ifvmode\nobreak\fi\fi\fi\@esphack}
             \gdef\@eqnlabel{#1}}
\def\@eqnlabel{}
\def\@vacuum{}
\def\draftmarginnote#1{\marginpar{\raggedright\scriptsize\tt#1}}

\def\draftlabel#1{{\@bsphack\if@filesw {\let\thepage\relax
        \xdef\@gtempa{\write\@auxout{\string
           \newlabel{#1}{{\@currentlabel}{\thepage}}}}}\@gtempa
        \if@nobreak \ifvmode\nobreak\fi\fi\fi\@esphack}
             \gdef\@eqnlabel{#1}}
\def\@eqnlabel{}
\def\@vacuum{}
\def\draftmarginnote#1{\marginpar{\raggedright\scriptsize\tt#1}}

\def\draft{\oddsidemargin -.5truein
             \def\@oddfoot{\sl preliminary draft \hfil
             \rm\thepage\hfil\sl\today\quad\militarytime}
             \let\@evenfoot\@oddfoot \overfullrule 3pt
             \let\label=\draftlabel
             \let\marginnote=\draftmarginnote
        \def\@eqnnum{(\theequation)\rlap{\kern\marginparsep\tt\@eqnlabel}%
\global\let\@eqnlabel\@vacuum}  }

\def\numberbysection{\@addtoreset{equation}{section}
             \def\theequation{\thesection.\arabic{equation}}}

\def\underline#1{\relax\ifmmode\@@underline#1\else
             $\@@underline{\hbox{#1}}$\relax\fi}

\def\titlepage{\@restonecolfalse\if@twocolumn\@restonecoltrue\onecolumn
          \else \newpage \fi \thispagestyle{empty}\c@page\z@
             \def\thefootnote{\fnsymbol{footnote}} }

\def\endtitlepage{\if@restonecol\twocolumn \else  \fi
             \def\thefootnote{\arabic{footnote}}
             \setcounter{footnote}{0}}  
\relax

\makeatletter
\newdimen\normalarrayskip              
\newdimen\minarrayskip                 
\normalarrayskip\baselineskip
\minarrayskip\jot
\newif\ifold             \oldtrue            \def\new{\oldfalse}
\def\arraymode{\ifold\relax\else\displaystyle\fi} 
\def\eqnumphantom{\phantom{(\theequation)}}     
\def\@arrayskip{\ifold\baselineskip\z@\lineskip\z@
         \else
         \baselineskip\minarrayskip\lineskip2\minarrayskip\fi}
\def\@arrayclassz{\ifcase \@lastchclass \@acolampacol \or
\@ampacol \or \or \or \@addamp \or
       \@acolampacol \or \@firstampfalse \@acol \fi
\edef\@preamble{\@preamble
      \ifcase \@chnum
         \hfil$\relax\arraymode\@sharp$\hfil
         \or $\relax\arraymode\@sharp$\hfil
         \or \hfil$\relax\arraymode\@sharp$\fi}}
\def\@array[#1]#2{\setbox\@arstrutbox=\hbox{\vrule
         height\arraystretch \ht\strutbox
         depth\arraystretch \dp\strutbox
         width\z@}\@mkpream{#2}\edef\@preamble{\halign
\noexpand\@halignto
\bgroup \tabskip\z@ \@arstrut \@preamble \tabskip\z@ \cr}%
\let\@startpbox\@@startpbox \let\@endpbox\@@endpbox
      \if #1t\vtop \else \if#1b\vbox \else \vcenter \fi\fi
      \bgroup \let\par\relax
      \let\@sharp##\let\protect\relax
      \@arrayskip\@preamble}
\def\eqnarray{\stepcounter{equation}%
                  \let\@currentlabel=\theequation
                  \global\@eqnswtrue
                  \global\@eqcnt\z@
                  \tabskip\@centering
                  \let\\=\@eqncr
     \halign to \displaywidth\bgroup
        \eqnumphantom\@eqnsel\hskip\@centering
        $\displaystyle \tabskip\z@ {##}$%
        \global\@eqcnt\@ne \hskip 2\arraycolsep
             $\displaystyle\arraymode{##}$\hfil
        \global\@eqcnt\tw@ \hskip 2\arraycolsep
             $\displaystyle\tabskip\z@{##}$\hfil
             \tabskip\@centering
        &{##}\tabskip\z@\cr}
\begingroup\ifx\undefined\newsymbol \else\def\input#1 {\endgroup}\fi
\newfont{\hr}{msbm10}
\newfont{\ams}{msam10}

\def\beq{\begin{equation}}
\def\eeq{\end{equation}}
\def\ba{\beq\new\begin{array}{c}}
\def\ea{\end{array}\eeq}

\hybrid

\def\beq{\begin{equation}}
\def\eeq{\end{equation}}
\def\p{\partial}

\begin{document}

\begin{titlepage}

\title{New applications of non-hermitian random
matrices\footnote{Talk given at TH-2002, Paris, UNESCO,
July 22-27 2002}}

\author{A.Zabrodin
\thanks{Institute of Biochemical Physics,
Kosygina str. 4, 119991 Moscow, Russia
and ITEP, Bol. Cheremushkinskaya str. 25, 117259 Moscow, Russia}}

\date{October 2002}
\maketitle



\begin{abstract}

We discuss recently discovered links of the statistical
models of normal random matrices to some
important physical problems of pattern formation and to
the quantum Hall effect. Specifically, the large $N$ limit of
the normal matrix model with a general statistical weight
describes dynamics of the interface between two incompressible
fluids with different viscousities in a thin plane cell
(the Saffman-Taylor problem). The latter appears to be
mathematically equivalent to the growth of semiclassical
2D electronic droplets in a strong uniform magnetic field
with localized magnetic impurities (fluxes), as the number of
electrons increases. The equivalence is most easily seen
by relating the both problems to the matrix model.

\end{abstract}

\vfill
ITEP-TH-46/02

\end{titlepage}

\section{Introduction}

The subject of the theory of random matrices is
a random matrix ${\sf M}$ distributed with some probability
measure $d\mu ({\sf M})$.
Typically, one is interested
in the distribution of eigenvalues and correlations
between them as size of the matrix, $N$, tends to infinity.

The range of physical applications of this theory
is enormously vast, with the role played by the
matrix ${\sf M}$ being very different.
In complex systems or systems with disorder
${\sf M}$ turns out to be a good substitute for the Hamiltonian or
transfer matrix.
Physical characteristics of the system are obtained via
averaging over one or another ensemble of large
matrices. Most extensively employed
and well-studied are ensembles of hermitian matrices.
In more recent applications
to statistical models on random lattices
and to string theory the key element is a set of graphs
in the diagrammatic expansion of random matrix integrals
while ${\sf M}$ has no physical meaning by itself.
For different aspects of random matrix theory and
related topics see e.g. \cite{Mehta,review1,review2}.

\begin{figure}[tp]
\epsfysize=5cm
\centerline{\epsfbox{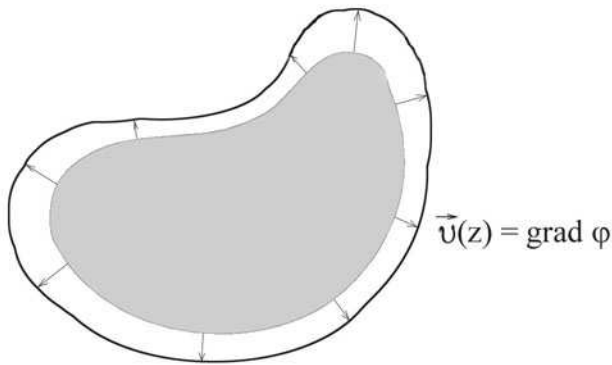}}
\caption{\sl Growth of the support of eigenvalues.}
\label{fi:growth}
\end{figure}

Complex non-hermitian random matrices are
employed in physics too. (A list of the relevant
physical problems and corresponding references
can be found in, e.g., \cite{list}.)
New applications
we are going to discuss are related to the distribution
of their eigenvalues.
To be specific, we consider the model of normal
random matrices, i.e., such that ${\sf M}$ commutes with
its hermitian conjugate, though similar results may hold
for other ensembles. Eigenvalues of normal matrices
are in general complex numbers.
When $N$ becomes large, they densely fill a domain
in the complex plane, the support of eigenvalues, with the mean
density outside it being zero. The shape of this domain is
determined by the probability measure and by the size of the
matrix. As $N$ increases,
the domain grows (see Fig.~\ref{fi:growth}).
The growth law is our main concern in this paper.

For simplicity we assume that the support of eigenvalues, $D$,
is a connected domain.
Let the size of the matrix grow linearly in time $t$:
$N \sim t$.
Then the support
of eigenvalues grows in such a way that the normal velocity of the
boundary is
\beq\label{intr1}
\vec v(z)=\mbox{grad}\, \varphi (z)\,,
\;\;\;\;\;\; z\in \p D
\eeq
where $\varphi$ is a function such that
\beq\label{intr2}
\left \{
\begin{array}{ll}
\Delta \varphi (z)=0 &\;\; z\in {\bf C}\setminus D
\\&\\
\varphi (z) \sim \log |z| &\;\; z\to \infty
\\&\\
\varphi (z) =0 &\;\; z\in \p D
\end{array}
\right.
\eeq
Here $\Delta =\p_{x}^{2}+\p_{y}^{2} =4\p_z \p_{\bar z}$
is the Laplace operator.
We employ the complex notation $z=x+iy$.
So, the dynamics of the boundary is governed
by the function
$\varphi$ which is harmonic in the exterior
of $D$ with a source at infinity, and
vanishes on the boundary.
The solution of the boundary
problem (\ref{intr2}) is unique:
$\varphi (z)=\log |w(z)|$,
where $w(z)$ is the conformal map from the exterior
of the domain $D$ onto the exterior of the unit circle
such that $\infty$ is mapped to $\infty$.
Such a map exists by virtue of the Riemann mapping theorem.

This dynamics is realized in various physical problems,
classical and quantum.
First, it is pertinent to viscous flows
and other pattern formation processes
when the normal velocity of the moving front
is proportional to the gradient of a harmonic field (see,
e.g., \cite{RMP}).
This mechanism is known as
Darcy's law. The droplet of eigenvalues behaves like
an incompressible
fluid with negligible viscosity (say water)
surrounded by a viscous fluid
(say oil), two fluids being confined in a thin plane
gap (the Hele-Shaw cell). Oil is withdrawn at infinity
at a constant rate while water is injected.
In this context, the function $\varphi$ is identified with the
pressure $P$ in the viscous fluid with
the opposite sign: $\varphi =-P$.
In water, the pressure can be set to $0$.
The condition that $P=0$ on the interface amounts
to neglecting the surface tension effects.
This idealization is good until the curvature
of the interface becomes large.
When the surface tension is
small enough, the dynamics becomes unstable.
The moving interface develops many fingers,
they split into new ones, and in a
sufficiently long time the water droplet
looks like a fractal.
In the literature, this phenomenon is refered to as
the Saffman-Taylor fingering.
This growth process was linked to
the matrix model in \cite{KKMWZ}.

As was recently pointed out in
\cite{ABWZ}, the same growth law applies to semiclassical dynamics
of an electronic droplet confined in a plane
on the lowest Landay level of a
strong magnetic field. This suggests
applications to the Quantum Hall effect.
It turns out that the shape of the electronic droplet
is sensitive to magnetic fluxes localized well away
from it. As one changes degeneracy of the level to
increase the number of electrons in the droplet, its shape
evolves in accordance with the Darcy law, thus showing up
fingering instabilities.
This phenomenon is purely quantum. Like the Aharonov-Bohm
effect, it is caused by quantum interference.
The function $\varphi$ in this case has no obvious physical
interpretation.
The characteristic scale of this phenomenon is
less than that of the Saffman-Taylor
fingering by a factor of $10^{9}$.
Remarkably, the matrix model provides
a unified mathematical treatment of the both phenomena.

From mathematical side, it is also worth noting
that calculation of certain expectation
values and correlation functions of normal random
matrices provides a constructive proof of some
important mathematical statements in the inverse
potential problem and the Dirichlet boundary problem
proved in a different way in \cite{MWZ}.

At last, the normal matrix model is known to be
integrable. Its partition function is a tau-function
of an integrable hierarchy of partial differential
equations. Although we do not discuss
integrability matters in this paper,
let us point out that
the above physical problems thus possess
a hidden integrable structure.

\section{Partition function of normal random matrices}

A matrix ${\sf M}$ is called normal if
$[{\sf M},\, {\sf M}^{\dag}]=0$, so that
${\sf M}$ and ${\sf M}^{\dag}$ can be simultaneously
diagonalized.
The model of normal random matrices was introduced
in \cite{normal}. The integrable structure of the model
is quite similar to that of the more widely known
model of two hermitian random matrices \cite{herm}
but physical interpretation is very different.

The partition function is
\beq\label{E1}
Z_N =\int_{\rm normal} d\mu_0 ({\sf M})
e^{\frac{1}{\hbar}{\rm tr} \, V({\sf M}, {\sf M}^{\dag})}
\eeq
where $\hbar$ is a parameter and $V$ is a real-valued
function.
The measure $d\mu_0$
is induced from the standard flat metric on the space
of all complex matrices.
To introduce coordinates in the subspace of normal matrices,
one makes use of the decomposition
${\sf M}={\sf U}{\sf Z}{\sf U}^{\dag}$ of a normal matrix
${\sf M}$, where
${\sf U}$ is a unitary matrix and ${\sf Z}=\mbox{diag}\,(z_1,
\ldots , z_N )$ is the diagonal matrix of eigenvalues
of the ${\sf M}$.
The measure is then given by
$$
d\mu_0 ({\sf M})=\frac{d\mu_{{\cal U}(N)}({\sf U})}{N! \,
\mbox{Vol}\, ({\cal U}(N))}
|\Delta_{N}(z)|^2 \prod_{i=1}^{N}d^2z_i
$$
where  $d\mu_{{\cal U}(N)}$ is
the Haar measure on the unitary group ${\cal U}(N)$, and
$\Delta_N(z)=
\prod_{i>j}^{N}(z_i -z_j)$
is the Vandermonde determinant.
The partition function is, therefore, written
as the following integral over eigenvalues:
\beq\label{E2}
Z_N = \frac{1}{N!}\int |\Delta_{N}(z)|^2
\prod_{j=1}^{N}\left ( e^{\frac{1}{\hbar}V(z_j )}
d^2z_j \right )
\eeq
(For notational simplicity we shall write $V(z)$ instead of
$V(z, \bar z)$.)
This quantity has two important interpretations.

One of them is the Coulomb gas picture \cite{Dyson}.
Writing
$Z_N =$ $\displaystyle{\frac{1}{N!}\int e^{{\cal E} (z_i)}
\prod_{j=1}^{N}d^2z_j}$
where
\beq\label{gas}
{\cal E}=\underbrace{\sum_{i\neq j}
\log |z_i -z_j |}_{\mbox{2D Coulomb energy}}\!\! +\,
\hbar^{-1}\underbrace{\sum_i V(z_i )}_{\mbox{potential}}
\eeq
we see that $Z_N$
is the partition function of the 2D Coulomb gas in the
external potential. Another one is the Quantum Hall
picture suggested in \cite{ABWZ}.

\paragraph{Quantum Hall picture.}
Consider spin-$\frac{1}{2}$ electrons on the plane
in a non-uniform magnetic field
$B$. The Pauli hamiltonian is
$$
H=\frac{1}{2m}\left ( (-i\hbar \nabla -\vec A)^2
-\hbar \sigma_3 B\right )
$$
where $\vec A$ is the vector potential.
If the magnetic field is uniform, the spectrum consists
of equidistant Landau levels, each level being highly
degenerate. The lowest level is very special.
Due to a hidden supersymmetry of the problem, it
can be found exactly and remains highly degenerate
even for arbitrary non-uniform field $B$ \cite{ACAS}.
The energy of this level equals $0$ while
the degeneracy equals the integer part of the total
magnetic flux $\Phi =\int B(z) d^2z$ in units of the flux quantum
$\Phi_0 =2\pi \hbar$ (we set $e=c=1$).
One-particle states on the lowest level can be found
explicitly. In the gauge
$A_x =\frac{1}{2}\p_y V$,
$A_y =-\frac{1}{2}\p_x V$ they are
$$
\psi_n (z)=P_n (z) \exp \left ( \frac{V(z)}{2\hbar}\right )
$$
where $B(z)=-\frac{1}{2}\Delta V(z)$.
Here $P_n =z^n + [\mbox{terms of lower degree}]$
are holomorphic
polynomials of any degree which is less than the degeneracy
of the level \cite{ACAS}.

Neglecting interactions between electrons,
the wave function of $N$ particles on the
lowest level is the Jastrow determinant:
$\Psi_N \sim  \frac{1}{\sqrt{N!}}
\det \psi_n (z_m)$,
and so
\beq\label{H1}
|\Psi_N |^2 = \frac{1}{N!} |\Delta_N (z_i)|^2 e^{\frac{1}{\hbar}
\sum_n V(z_n  )}
\eeq
coincides with the statistical weight of normal random matrices
expressed through eigenvalues.
The partition function (\ref{E2}) is, in this context, the
normalization factor of the $N$-particle wave function:
$\int |\Psi_N|^2 \prod_i d^2z_i = Z_N$.
The mean density of electrons
coincides with the expectation
value of the density of eigenvalues in the matrix model:
$$
\displaystyle{NZ_{N}^{-1}
\int }|\Psi_N(z, \xi_1 , \ldots , \xi_{N-1})|^2
\prod_{i=1}^{N-1}d^2 \xi_i
=\left <\mbox{tr}\, \delta^{(2)}(z-{\sf M})\right >
$$
Similarly, multiparticle correlation functions
are identified with multipoint correlation functions
of densities.

All the above relations are exact at any finite $N$.
As $\hbar$ becomes small and $N$ large, one approaches
a semiclassical regime. However, the semiclassical properties
of the system are quite
unusual. On the one hand, the density distribution
acquires a well-defined edge, and one can speak about a
well localized electronic
droplet which behaves like an
incompressible fluid. On the other hand, in this
specific semiclassical regime, quantum effects are
by no means negligible. Quite the reverse, they
become rather strong if not dominant.
In fact there is no surprize here because the
semiclassical limit we are speaking about is not
the usual one which would require excitations of higher
energy levels. In our ``semiclassical" limit all
particles occupy the lowest level, so the
droplet as a whole remains a quantum object.
Amusingly enough, it is this limit where one makes
contact with the purely classical
Saffman-Taylor fingering.
In the next section, we analyse the corresponding
large $N$ limit
of the matrix integral.

\section{The semiclassical (large $N$) limit}

The large $N$ limit we are interested in is
$N \to 0$, $\hbar \to 0$ with $\hbar N$ finite and fixed.
The expansion
in $N^{-1}$ is then the same as the expansion in $\hbar$.

To elaborate the limit,
we represent the energy ${\cal E}$ (\ref{gas}) in the form
$$
{\cal E}=\int \!\int \rho (z)\log |z-z'|\rho (z')
d^2z d^2 z' +\frac{1}{\hbar}\int V(z)\rho (z) d^2z
$$
where $\rho (z)=\sum_i \delta^{(2)}(z-z_i)$ is the density
of eigenvalues.
In the limit, one treats $\rho (z)$
as a continuous function normalized as $\int \rho (z)d^2z=N$.
As $\hbar \to 0$, both terms in
(\ref{gas}) are of order $N^2$, and the saddle point method can be
applied to perform the integral.

The saddle point condition is
$\delta {\cal E}/\delta \rho (z) =0$ which yields
the integral equation for the mean density:
\beq\label{semitau1}
\hbar \int \frac{\rho (z')d^2z'}{z-z'}
+\p_z V(z) =0
\eeq
The meaning of this equation is especially clear
in the Coulomb gas interpretation. It states that
each charge is in the equilibrium. Indeed,
consider a charge at the point $z$. The first term
in the equation
is the Coulomb force caused by other charges in the gas
while the second term is the external force. The equation
just tells that they compensate each other.
Clearly, it makes sense to impose the
equilibrium condition only
in the domain where the charges are actually present,
i.e., in the support of eigenvalues.

So, the equation should be
satisfied in a domain $D$ where $\rho \neq 0$.
Here we assume that $D$ is a connected domain.
For example, in the potential $V=-|z|^2$ the eigenvalues
uniformly fill the disk of radius $\sqrt{\hbar N}$.
Small perturbations of the potential
slightly disturb the circular shape.

It appears that in case of
normal matrices the above integral equation is much easier
to solve than the similar
equation for distribution of eigenvalues of hermitian matrices.
Indeed, on applying $\p_{\bar z}$
to both sides of eq.\,(\ref{semitau1}) we obtain
$\rho (z)=- \frac{1}{4\pi \hbar}\Delta V (z)$
in $D$,
and $\rho (z)=0$ in ${\bf C}\setminus D$.
The domain $D$ itself is determined by the condition
\beq\label{saddle1}
\oint_{\p D}\frac{\p_{z'}V(z') dz'}{z-z'} =0
\eeq
which can be derived from (\ref{semitau1})
with the help of the Cauchy integral formula.
The condition means, in other words, that $D$ is
such that the function $\p_z V$ on its boundary is the
boundary value of an analytic function in
${\bf C}\setminus D$.

An important particular case is $V$ equal to
$-|z|^2$ plus a harmonic function which we parametrize
by its Taylor coefficients $t_k$ at the origin:
$V=-|z|^2 +2{\cal R}e \,\sum t_k z^k$.
Then the density is constant (equal to $\frac{1}{\pi \hbar}$)
inside $D$ and zero outside. The area of $D$ is equal to
$\hbar N$.
The shape of $D$ is determined by the conditions
$$
-\frac{1}{\pi k}
\int_{{\bf C}\setminus D}z^{-k}d^2z =t_k
$$
which easily follow from (\ref{saddle1}),
so that $\pi k t_k$ are harmonic moments of the
domain complementary to $D$.
In case of quadratic potential the domain is
an ellipse \cite{FGIL}.

The integrated version of eq.\,(\ref{semitau1}) allows one
to find the leading contribution to the free energy, which is
given by the value of $\hbar^2 {\cal E}$ at the saddle point.
Let us denote the latter by $F_0$, then
\beq\label{F0}
F_0= -\, \frac{1}{16\pi^2}
\int_D \!\! \int_D \Delta V (z)\log \left |
\frac{1}{z}-\frac{1}{z'}\right | \Delta V(z')
d^2z d^2z'
\eeq
This is the tau-function of curves introduced in
\cite{KKMWZ}.
The leading asymptotics of the partition function as
$\hbar \to 0$ is therefore $Z_N \simeq e^{F_0 /\hbar^2}$.

\begin{figure}[tp]
\epsfysize=5cm
\centerline{\epsfbox{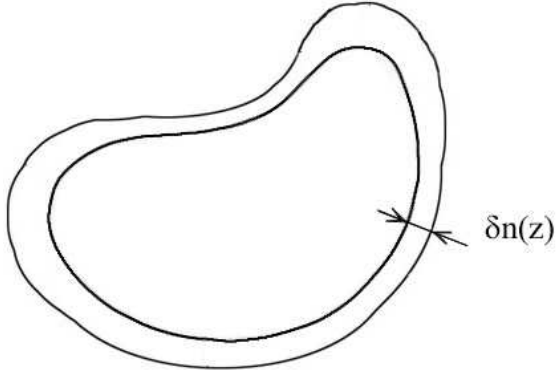}}
\caption{\sl Normal displaycement of the contour.}
\label{fi:deltan}
\end{figure}

\paragraph{Small variations of the potential
and the growth law.}
If one varies the potential, $V\to V+\delta V$, and
size of the matrix, $N \to N+\delta N$,
the support of
eigenvalues slightly changes its shape and area.
Let us examine how it goes. It is natural to characterize
an infinitesimal change of the boundary by its normal displaycement
$\delta n(z)$ at each point $z$,
so that $\delta n(z)$ is a continuous function on the boundary curve
(see Fig.~\ref{fi:deltan}).

First we vary the potential at constant $N$.
The shape of the support of eigenvalues is
determined by eq.\,(\ref{saddle1}). Its variation
can be written as
\beq\label{var11}
\oint \frac{\p_{z'}\delta V(z') dz'}{z-z'}
+\frac{i}{2}\oint \frac{\Delta V(z')
\delta n(z')}{z-z'} |dz'| =0
\eeq
It is natural to employ the ansatz
$\Delta V (z)\delta n(z)=\p_n h(z)$ where $h$ is yet
unknown function in the exterior of $D$ such that
$h=0$ on the boundary, and $\p_n$ means its normal
derivative, with the normal vector pointing outward.
This ansatz is suggested
by an easy transform of the second integral into
a Cauchy integral. Combining the two terms, we get
$\oint \frac{\p_{z'}(\delta V(z')+h(z'))}{z-z'}
\, dz' =0$ for  $z\in D$.
This implies that the function $\p_z (\delta V +h)$
is analytic in the exterior of $D$, i.e., that the function
$\delta V +h$ is harmonic in there. Hence $h(z)=
\delta V^H (z) -\delta V$ where, given a function
$f$, we use the notation
$f^H$ for the harmonic continuation of this function
from the boundary to the exterior of $D$.
Therefore,
\beq\label{CC4}
\delta n(z)=\frac{\p_n \left ( \delta V^H(z)-
\delta V(z)\right )}{\Delta V(z)}
\eeq

Similarly, to find how the domain $D$ grows
at constant $V$ as $N$ increases,
we use eq.\,(\ref{saddle1}) again. This time
the first term in (\ref{var11}) is zero.
Noting that
$$
\delta N =-\, \frac{1}{4\pi \hbar} \oint \delta n(z)
\Delta V (z) |dz|
$$
it is easy to see that the solution is
\beq\label{var1}
\delta n(z) =-\, \frac{2\hbar \delta N}{\Delta V (z)}
\p_n \log |w(z)|
\eeq
where $w(z)$ is the conformal map from ${\bf C}\setminus D$
onto the exterior of the unit circle such that $\infty$
is sent to $\infty$.
Since $\log |w| =0$ on the boundary, the normal derivative
can be substituted by gradient. Therefore,
at $\Delta V =\mbox{const}$ we get the Darcy law
(\ref{intr1}) (cf.\,\cite{KKMWZ}).

\paragraph{Semiclassical electronic droplet in the presence
of magnetic impurities.} Let us apply the above results
to the semiclassical behaviour of an electronic droplet
in a strong magnetic field.
The notion of the Quantum Hall droplet \cite{QH}
implies that the electronic
liquid is incompressible, i.e., all states at the
lowest energy level are occupied. Therefore, we
want the degeneracy of the level to be equal to $N$.
This can be achieved in different ways. One of them
is to assume the following arrangement.
Let a strong uniform magnetic field $B_0 >0$ be applied
in a large disk of radius $R_0$. The disk is surrounded
by a large annulus $R_0 <|z|<R_1$ with a magnetic field
$B_1 <0$ such that the total magnetic flux through the system
is $N\Phi_0$. The magnetic field outside the largest disk
$|z|<R_1$ vanishes. The disk is connected through a tunnel
barier to a large capacitor that maintains a small positive
chemical potential slightly above the zero energy.
If $B_0$ is strong enough, the gap
is large, and the higher levels can be neglected.
In this arrangement, the circular droplet of $N$ electrons
is trapped at the center. Its radius is much less
than $R_0$.
The function $V(z)$ for $|z|<R_0$ is
$V(z)=-\frac{1}{2}B_0 |z|^2$.

Now let us apply a non-uniform magnetic field $\delta B$
somewhere
inside the disk $|z|<R_0$ but well away from the droplet.
Suppose that the nonuniform magnetic field
does not change the total flux: $\int \delta B d^2z =0$.
The potential $V(z)$ inside and around the droplet
is modified as
\beq\label{semi}
V(z)=-\frac{B_0}{2}|z|^2 -
\frac{1}{\pi}\int \log |z-z'|\delta B(z')d^2z'
\eeq
The second term is harmonic inside and around the droplet.
One may have in mind thin solenoids
carrying magnetic flux (``magnetic impurities").
In the case of point-like magnetic fluxes
$q_i$ at points $a_i$, we have
$V(z)=-\frac{1}{2}B_0 |z|^2 +
\sum_i q_i \log |z-a_i |$.

\begin{figure}[tp]
\epsfysize=5cm
\centerline{\epsfbox{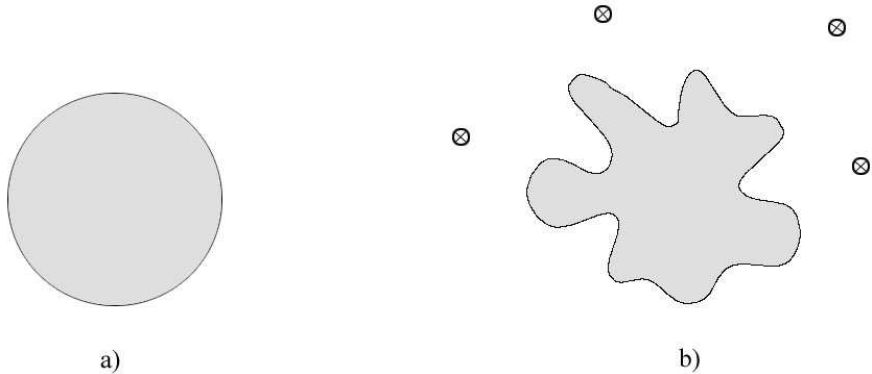}}
\caption{\sl a) The electronic droplet in the uniform magnetic field;
b) The electronic droplet in the uniform magnetic field in the
peresence of point-like fluxes at the points
marked by $\otimes$.}
\label{fi:ameba}
\end{figure}

In the presence of the fluxes,
the shape of the droplet is no longer circular
(Fig.~\ref{fi:ameba})
although the magnetic field inside the droplet
and not far from it
remains uniform and is not changed at all. In this respect
this phenomenon is similar to the Aharonov-Bohm
effect. The responce of the droplet to an
infinitesimal change of the magnetic field
$\delta B$ is described by eq.\,(\ref{CC4})
in which
$$
\delta V^H(z) -\delta V(z)=
\frac{1}{\pi}\int_{{\bf C}\setminus D}
G(z,z')\delta B(z') d^2z'
$$
Here $G(z,z')$ is the Green function of the
Dirichlet boundary problem in ${\bf C}\setminus D$
normalized in such a way that $G(z,z')\to \log |z-z'|$
as $z\to z'$. In fact this formula holds for arbitrary
$\delta B$, not necessarily vanishing inside the droplet.
In particular, for small point-like fluxes
$\delta q_i$ at some points $a_i$ we have
$\delta V =\sum_i \delta q_i \log |z-a_i|$,
$\delta B=-\pi \sum_i \delta q_i \delta^{(2)}(z-a_i)$, and
$$
\delta V^H(z) -\delta V(z)=
-\sum_i  \, G(z, a_i)\delta q_i
$$
If $a_i$ is inside, $G(z, a_i)$ is set to be zero.
The sum, therefore, goes over outside fluxes only.
The fluxes inside the droplet, if any,
appear to be completely screened and do not have any influence
on its shape.

When $B_1$ adiabatically increases, with
$B_0$ and $\delta B$ fixed, the droplet grows because
the degeneracy of the lowest level is enlarged and new
electrons enter the system.
The growth is described by eq.\,(\ref{var1})
with $\Delta V (z)=- 2 B_0$ which is
equivalent to the Darcy law.

\section{Conclusion}

We have analysed the large $N$ limit of the model
of normal random matrices.
It has been argued that as $N$ increases,
the growth of the support of
complex eigenvalues simulates important physical
phenomena:
\begin{itemize}
\item
Interface dynamics in viscous flows (the Saffman-Taylor
fingering) in the zero surface tension limit
\item
Semiclassical behaviour of 2D electronic droplets
in the Quantum Hall regime
\end{itemize}
The former is purely classical while the latter is
purely quantum.

The relation to the matrix model may help
to suggest a way to regularize singularities
which usually occur in the zero surface tension limit
and to obtain an analytically tractable formulation
of the Saffman-Taylor problem with surface tension.

\section*{Acknowledgments}
I am grateful to O.Agam, E.Bet\-tel\-heim, I.Kos\-tov,
I.Kri\-che\-ver, A.Mar\-sha\-kov, M.Mi\-ne\-ev-\-Wein\-stein
and P.Wieg\-mann for collaboration.
This work was supported in part
by the LDRD project 20020006ER ``Unstable
Fluid/Fluid Interfaces" at Los Alamos National Laboratory
during the author's visit at LANL in January-February 2002,
by RFBR grant 00-02-16477,
and by grant INTAS-99-0590.

\end{document}